\documentclass[twocolumn,showpacs,preprintnumbers,amsmath,amssymb,prl]{revtex4}
\usepackage{graphicx}
\begin{document}
\title{Realization of four-state qudits using biphotons}
\author {E. V. Moreva}
\email{ekaterina.moreva@gmail.com} \affiliation{Moscow Engineering
Physics Institute (State University) Russia}
\author{G. A. Maslennikov}
\affiliation{National University of Singapore, Singapore}
\author{S. S. Straupe, and S. P. Kulik}
\affiliation{Faculty of Physics, Moscow State University, Russia}

\date{\today}
\begin{abstract}
The novel experimental realization of four-level optical quantum
systems (ququarts) is presented. We exploit the polarization
properties of the frequency non-degenerate biphoton field to obtain
such systems. A simple method that does not rely on interferometer
is used to generate and measure the sequence of states that can be
used in quantum key distribution (QKD) protocol.
\end{abstract}
\pacs{42.50.-p, 42.50.Dv, 03.67.-a} \maketitle \textbf{I.
Introduction.} Recently multi-dimensional (D$>$2) quantum systems or
qudits attracted much attention in context of quantum information
and communication. It is partly caused by fundamental aspects of
quantum theory since the usage of qudits allows one to violate
Bell-type inequalities longer than with two dimensional systems (for
the references see the review \cite{quBell}). Much interest in
qudits also comes from the application point, especially from the
applied quantum key distribution (QKD). Multilevel systems are
proved to be more robust against noise in the transmission channel,
although measurement and preparation procedures of such systems
seems to be much more technically complicated than in the case of
qubits. Different aspects of the security of qudit-based protocols
have been analyzed \cite{Security,Tittel}. Lately a
proof-of-principle realization of a QKD protocol with entangled
qutrits ($D=3$) \cite{QKDZeilinger:05} and with qudits
\cite{Walborn:05} have been demonstrated. For the last years several
elegant experiments have been performed where different kinds of
optical qudits were introduced
\cite{Walborn:05,Howell:05,OAM,timebin,Thew:04,rob:01,antia:01,ourPRL:04}.
Recently an experiment that ensured the full control over a
polarization qutrit state was demonstrated
\cite{ourPRL:04,ourPRA:04}. However polarization qutrits do not seem
to serve as a practical candidate for the multilevel QKD, since it
is impossible to achieve a demanded state using only SU2
transformations that are done with linear optical elements. Meantime
transformations done by these elements are needed for practical
realization of QKD protocols. \\In this Letter we present the
results of the experimental preparation, transformation, and
measurement carried out with polarization based ququarts or quantum
systems with dimensionality $D=4$.\\\textbf{II. Polarization
ququarts and their properties.} If one considers the two-photon
field, generated via spontaneous parametric down-conversion (SPDC)
process, then the pure polarization state can be written as a
superposition of four basic components: \small
\begin{equation}
|\Psi\rangle=c_{1}|H_{{1}},H_{{2}}\rangle+c_{2}|H_{{1}},V_{{2}}\rangle+
c_{3}|V_{{1}},H_{{2}}\rangle+c_{4}|V_{{1}},V_{{2}}\rangle.
\label{eq:state}
\end{equation}
\normalsize Here $c_i=|c_i|e^{i\phi_i},(i=1,2,3,4)$ are complex
probability amplitudes, $|H_{{j}}\rangle\equiv
a_{\lambda_{j}}^{\dagger}|vac\rangle$, $|V_{{j}}\rangle\equiv
b_{\lambda_{j}}^{\dagger}|vac\rangle$, where $\lambda_{j},(j=1,2)$
are the central wavelengths of down converted photons. If the down
converted photons have an only polarization degree of freedom, then
a ququart state (\ref{eq:state}) converts to a qutrit state i.e.
middle terms in (\ref{eq:state}) become indistinguishable. In order
to distinguish between these terms one must be able to distinguish
between the down converted photons either in frequency, momentum or
detection time. In experiments, described in this paper, we chose
the collinear non-degenerate regime of SPDC, so twin photons that
form a biphoton were having different frequencies and propagating
simultaneously along the same direction. The sum of their
frequencies was equal to the frequency of the pump, according to
energy conservation. Polarization properties of this state can be
described by Stokes parameters, which are defined as the mean values
of Stokes operators, averaged over a state (\ref{eq:state}).
Although the description of the light polarization can be introduced
only for the quasimonochromatic plane waves, it is possible to use
$P$-quasispin formalism \cite{karasev} to describe the polarization
of arbitrary quantum beams with $n$ modes, frequency or spatial. It
is worthy to note that the frequency representation of the ququart
(\ref{eq:state}) is isomorphic to the spatial one, when twin photons
have the same frequencies but propagate in different directions
\cite{Kwiat:01}. Frequently in the tasks of quantum communication it
is convenient to operate with states in a single spatial mode. For
the two-frequency and single-spatial mode field, the formal
definition of annihilation/creation operators is given by the sum of
corresponding operators in each mode. Also we take into account that
these operators do commute for different frequency modes. So the
Stokes parameters will contain time dependent terms
$exp(i(\omega_{1}-\omega_{2})t)$ that describe "beats" of frequency
modes and have no connection with the light polarization. However,
these terms vanish if one considers the finite detection time, that
allows to classically average these "beats":

\begin{equation}
\begin{array}{cc}
\langle S_{0}\rangle=\langle
a_{1}^{\dagger}a_{1}+a_{2}^{\dagger}a_{2}+b_{1}^{\dagger}b_{1}+b_{2}^{\dagger}b_{2}\rangle=2;
&\\\langle S_{1}\rangle=\langle
a_{1}^{\dagger}a_{1}+a_{2}^{\dagger}a_{2}-b_{1}^{\dagger}b_{1}-b_{2}^{\dagger}b_{2}\rangle
= 2(|c_{1}|^2-|c_{4}|^2); &\\
\langle S_{2}\rangle=\langle
a_{1}^{\dagger}b_{1}+a_{2}^{\dagger}b_{2}+b_{1}^{\dagger}a_{1}+b_{2}^{\dagger}a_{2}\rangle
= \\ 2Re(c_{1}^{\star}(c_{2}+c_{3})+c_{4}(c_{2}^{\star}+c_{3}^{\star})); &\\
\langle S_{3}\rangle=\langle
a_{1}^{\dagger}b_{1}+a_{2}^{\dagger}b_{2}-b_{1}^{\dagger}a_{1}-b_{2}^{\dagger}a_{2}\rangle
=&\\2Im(c_{1}^{\star}(c_{2}+c_{3})+c_{4}(c_{2}^{\star}+c_{3}^{\star})).
\end{array}
\label{eq:Stokes}
\end{equation}
The polarization degree is given by
\begin{equation}
P_{4}={\frac{\sqrt{\sum\limits_{k=1}\limits^{3}{\langle
S_{k}^{(1)}+S_{k}^{(2)}\rangle}^2}}{\langle
S_{0}^{(1)}+S_{0}^{(2)}\rangle}}. \label{eq:poldeg}
\end{equation}
This definition of the polarization degree is just generalization of
the commonly used classical one. It differs from the definition
suggested in \cite{gunnar:05}, where it serves as a witness of the
state purity. In the case of polarization-based qutrit states
\cite{ourPRL:04, ourPRA:04}, the polarization degree $P_{3} = \sqrt
{|c_1^{\prime}|^2 - |c_3^{\prime}|^2 + 2| {c_1^{\prime\ast}
c_2^{\prime} + c_2^{\prime\ast} c_3^{\prime} } |^2} $ with $
c_1^{\prime}=c_1, \sqrt{2}c_2^{\prime}= c_2=c_3, c_3^{\prime}=c_4$
in (\ref{eq:state}) was an invariant to unitary polarization
transformations. Indeed it is impossible to prepare all demanded
pure states, unless one uses interferometric schemes with several
nonlinear crystals \cite{ourPRL:04}. In particular there is no way
to transform the basic qutrit state
$|\Psi_{4}^{\prime}\rangle=|V,V\rangle$ with $P$ = 1 into the state
$|\Psi_{2}^{\prime}\rangle=|H,V\rangle$ with $P$ = 0 using
retardation plates. However, in the case of polarization ququarts,
this quantity is no longer the invariant and can be changed by
applying local unitary transformations in each frequency mode. It
can be achieved by using dichroic polarization transformers, which
act separately on the photons with different frequencies. For
example to transform the state
$|\Psi_{4}\rangle=|V_{\lambda_{1}},V_{\lambda_{2}}\rangle$ into the
state $|\Psi_{2}\rangle=|H_{\lambda_{1}},V_{\lambda_{2}}\rangle$ one
needs to use the retardation plate which serves as a half wave plate
at $\lambda_{1}$ and as a wave plate at $\lambda_{2}$. The unitary
transformation on the state (\ref{eq:state}) is given by $4\times 4$
matrix that is obtained by a direct product of two $2\times 2$
matrices describing the transformation performed on each photon:
\begin{equation}
G \equiv \left( {{\begin{array}{*{20}c}
 {t_{1}t_{2}} \hfill & {t_{1}r_{2}} \hfill & {r_{1}t_{2}} \hfill & {r_{1}r_{2}} \hfill \\
 {-t_{1}r_{2}^*} \hfill & {t_{1}t_{2}^*} \hfill & {-r_{1}r_{2}^*} \hfill & {r_{1}t_{2}^*} \hfill \\
 {-r_{1}^*t_{2}} \hfill & {-r_{1}^*r_{2}} \hfill & {t_{1}^*t_{2}} \hfill & {t_{1}^*r_{2}} \hfill \\
 {r_{1}^*r_{2}^*} \hfill & {-r_{1}^*t_{2}^*} \hfill & {-t_{1}^*r_{2}^*} \hfill & {t_{1}^*t_{2}^*} \hfill \\
\end{array} }} \right),
\end{equation}
with "transmission" and "reflection" coefficients \\$t_{i}=\cos
\delta_{i} + i\sin \delta_{i} \cos 2\alpha_{i}, \quad r_{i} = i\sin
\delta_{i} \sin 2\alpha_{i},\\ \delta_{i} = {\pi (n_o - n_e) h/
\lambda_{i} }$, where $\delta_{i}$ is its optical thickness, $h$ is
the geometrical thickness, and $\alpha_{i}$ is the orientation angle
between the optical axis of a wave plate and vertical direction. So
for the unitary transformation of the ququart the matrix considered
above has the form
\begin{equation}
G\equiv \left( {{\begin{array}{*{20}c}
 {0} \hfill & {0} \hfill & {1} \hfill & {0} \hfill \\
 {0} \hfill & {0} \hfill & {0} \hfill & {1} \hfill \\
 {1} \hfill & {0} \hfill & {0} \hfill & {0} \hfill \\
 {0} \hfill & {1} \hfill & {0} \hfill & {0} \hfill \\
 \end{array} }} \right).
 \label{eq:platematrix}
\end{equation}
The same dichroic plate performs direct and reverse unitary
transformations between the polarization Bell states:
$|\Phi^{+(-)}\rangle$ and $|\Psi^{+(-)}\rangle$, which are
represented by ququarts with $c_{2}$=$c_{3}$=0,
$c_{1}$=+(-)$c_{4}$=$\frac{1}{\sqrt{2}}$ and $c_{1}$=$c_{4}$=0,\\
$c_{2}$=+(-)$c_{3}$=$\frac{1}{\sqrt{2}}$ correspondingly. Similar
transformations with frequency non-degenerate biphotons have been
realized in \cite{freqBelltrans}.

\par In general to prepare an arbitrary ququart state
(\ref{eq:state}) it is necessary to use four nonlinear crystals
arranged in such a way that each crystal emits coherently one basic
state in the same direction. But in particular cases a reduced set
of crystals is quite sufficient to generate specific ququart states
which can be used in applications. For example to prepare the
polarization Bell states it was sufficient to use two crystals
\cite{freqBelltrans, freqBellprep}. Moreover even single crystal
allows one to prepare the useful subset of ququarts.

\textbf{III. Polarization ququarts in QKD protocol.} The complete
QKD protocol with four-dimensional polarization states exploits five
mutually unbiased bases with four states in each. In terms of
biphoton states the first three bases consist of product
polarization states of two photons while the last two bases consist
of two-photon entangled states:
\begin{equation}
\begin{array}{cc}
\quad I.\quad|H_{1}H_{2}\rangle; \quad |H_{1}V_{2}\rangle;
\quad|V_{1}H_{2}\rangle; \quad|V_{1}V_{2}\rangle,\\
\quad II.\quad|D_{1}D_{2}\rangle; \quad|D_{1}{A_{2}}\rangle;\quad
|A_{1}D_{2}\rangle; \quad|A_{1}A_{2}\rangle, \\
III. \quad|R_{1}R_{2}\rangle; \quad|R_{1}L_{2}\rangle;\quad
|L_{1}R_{2}\rangle; \quad|L_{1}L_{2}\rangle,\\
IV. \quad|R_{1}H_{2}\rangle + |L_{1}V_{2}\rangle;\quad
|R_{1}H_{2}\rangle - |L_{1}V_{2}\rangle;\\
 \quad\quad|L_{1}H_{2}\rangle + |R_{1}V_{2}\rangle;\quad
|L_{1}H_{2}\rangle - |R_{1}V_{2}\rangle, \\
V. \quad|H_{1}R_{2}\rangle + |V_{1}L_{2}\rangle;\quad
|H_{1}R_{2}\rangle - |V_{1}L_{2}\rangle; \\
\quad\quad|H_{1}L_{2}\rangle + |V_{1}R_{2}\rangle;\quad
 |H_{1}L_{2}\rangle - |V_{1}R_{2}\rangle.
 \end{array}
 \label{qkd:states}
\end{equation}
Here $H,V,D,A,R,L$ indicate horizontal, vertical, +45 and -45
linear, right- and left-circular polarization modes correspondingly,
and lower indices numerate the frequency modes of two photons. It
has been proved \cite{Tittel} that using only first two or three
bases is sufficient for the efficient QKD. Exploiting the incomplete
set of bases one sacrifices the security but enhances the key
generation rate. That is why we will restrict ourselves to the first
three bases and skip the rest two. As we will show experimentally,
the set of twelve states can be prepared with a single non-linear
crystal and local unitary transformations. We also present a
measurement scheme that allows to discriminate the states belonging
to one basis deterministically thus allowing its implementation in
realization of QKD protocol with polarization ququarts.

\textbf{IVa. Experimental setup.} The experimental setup for
generation and measurement of ququart states is shown at

\begin{figure}[!ht]
\includegraphics[width=0.38\textwidth]{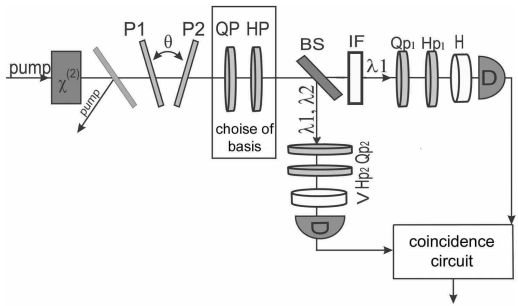}
\caption{Setup for preparation and measurement of ququarts}
\label{setup1}
\end{figure}

\begin{figure}[!ht]
\includegraphics[width=0.28\textwidth]{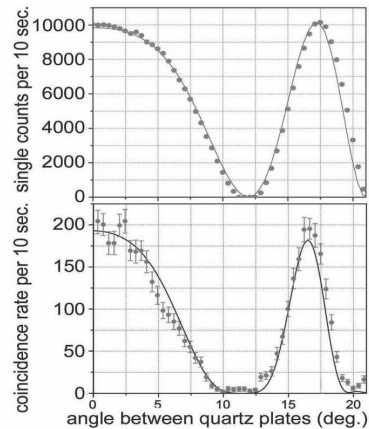}
\caption{Dependence of single counts (upper) and coincidences
(lower) on tilting angle ${\theta}$} \label{graphic}
\end{figure}

Fig.~\ref{setup1}.
The 10$mW$ cw He-Cd laser operating at 325$nm$ serves as a pump. A
15mm lithium-iodate type I crystal emits down-converted photons at
central wavelengths $\lambda_{1}=702 nm$,
 $\lambda_{2}=605nm$ within the spectral width of 2$nm$ each, propagating collinearly with the pump,
so the ququart state $|V_{1}V_{2}\rangle$ is generated.
 Then, this state was subjected to transformations done by dichroic
wave plate(s). Finally the state passed through the zero-order half-
and quarter plates depending on which bases has been chosen. The
measurement setup consists of a Brown-Twiss scheme with a
non-polarizing beam-splitter. An interference filter centered at
702$nm$ with a FWHM bandwidth of 3$nm$ was placed in transmitted
arm. This part of the scheme allows to reconstruct arbitrary
polarization biphoton-ququart by registering coincidence counts for
different projections that are done by the polarization filters
located in each arm \cite{JETPLett}. Each filter consists of a
zero-order quarter- and half-waveplate and a fixed analyzer. Two
Si-APD's, linked to a coincidence scheme with 1.5 nsec time window,
were used as single photon detectors.\\\textbf{IVb. Experimental
procedure.} Let us consider as an example, the preparation of a
state $|H_{1}V_{2}\rangle$ from the initial state
$|V_{1}V_{2}\rangle$. This transformation can be achieved by a
dichroic wave plate oriented at $45^\circ$ that introduces a phase
shift of $2\pi$ between extra- and ordinary polarized photons at
605$nm$, and a phase shift of $\pi$ for the conjugate photons at
702$nm$. Using quartz as birefringent material it is easy to
calculate that the one of possible thicknesses of the wave plate
that does this transformation should be equal to $3.406 mm$
\cite{order}. We used two plates P1 and P2 with an effective
thickness of 3.401$mm$. If then one can tilt these wave plates
towards each other by a finite angle $\theta$, then the optical
thickness of the effective wave plate will be changing and, at a
certain value of $\theta$, the desired transformation will be
achieved. This corresponds to maximal coincidence rate when the
measurement part is tuned to select $|H_{1}V_{2}\rangle$ state.
Monitoring the coincidences, one can obtain the value of ${\theta}$
for which the main maximum occurs. Then, fixing the tilting angle at
this value, one can perform a complete quantum state tomography
protocol in order to verify if the state really coincides with the
ideal. In order to change the basis from $I$ to $II$ ($III$), zero
order half- (quarter) wave plates oriented at $22.5^0$ ($45^0$) were
used.\\\textbf{V. Results and discussion.} Fig.~\ref{graphic}
shows the coincidences and single count rates versus the change of
tilting angle $\theta$ which determines optical thickness of the
effective wave plate. If the measurement setup is tuned to select
the state $|H_{1}V_{2}\rangle$ then dependence of coincidences rate
on the plate optical thicknesses $\delta_{i}$ is given by formula:
\begin{equation}
R_{coin}\propto\langle{{a_1}^\dagger{b_2}^\dagger{a}_1{b}_2}\rangle
=\sin^2(\delta_{1})\cos^2(\delta_{2}),
\end{equation}
whereas the single counts distribution in the upper channel is
given by
\begin{equation}
I_{702nm}\propto\langle{{a_1}^\dagger{a}_1}\rangle=\sin^2(\delta_{1}).
\label{singles}
\end{equation}
The solid lines at the Fig.~\ref{graphic} show the theoretical
curves. We performed tomography measurements for the main and
additional maxima as well as for the minimum. The minima in
coincidences occur when intensity in any channel drops to zero, so
it is not a necessary condition for distinguishing the orthogonal
state to the one selected by given settings of polarization filters.
Nevertheless according to calculations and our measurements the
minimum in the coincidences at Fig.~\ref{graphic} exactly refers to
the state $|V_{1}H_{2}\rangle$. Starting from the
$|V_{702nm}V_{605nm}\rangle$ we prepared and measured the whole set
of the states from (\ref{qkd:states}) belonging to the first three
bases. The table I show components of the experimental (theoretical)
density matrix as well as fidelity $F$ defined by $F = Tr(
\rho_{th}\rho_{exp})$ for some of the states.

\begin{table}[!ht]

\begin{tabular}{|c|c|c|c|c|c|}
\hline State & $\rho^{exp}_{11}(\rho^{th}_{11})$ & $\rho^{exp}_{22}(\rho^{th}_{22})$ & $\rho^{exp}_{33}(\rho^{th}_{33})$ & $\rho^{exp}_{44}(\rho^{th}_{44})$ & $F$ \\
\hline $|H_{1}V_{2}\rangle$ & 0(0) & 0.984(1) & 0(0) & 0.016(0) & 0.98 \\
\hline $|V_{1}H_{2}\rangle$ & 0.024(0) & 0.004(0) & 0.944(0) & 0.028(0) & 0.94 \\
\hline $|D_{1}A_{2}\rangle$ & 0(0) & 0.994(1) & 0(0) & 0.006(0) & 0.99 \\
\hline $|A_{1}D_{2}\rangle$ & 0.015(0) & 0(0) & 0.950(1) & 0.035(0) & 0.95 \\
\hline $|R_{1}L_{2}\rangle$ & 0(0) & 0.973(1) & 0.027(0) & 0(0) & 0.97 \\
\hline $|L_{1}R_{2}\rangle$ & 0.016(0) & 0(0) & 0.957(1) & 0.027(0) & 0.96 \\
\hline
\end{tabular}
\caption{Density matrix components and fidelities for selected
transformations.}
\end{table}

\par
\begin{figure}[!ht]
\includegraphics[width=0.3\textwidth]{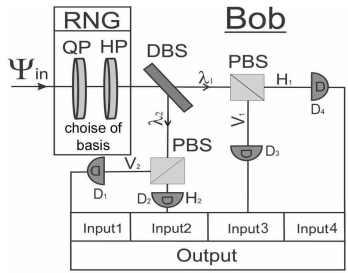}
\caption{Measurement part at Bob's station} \label{setup2}
\end{figure}

 The main obstacle for the practical implementation of the free-space QKD protocol
based on ququarts is that one needs to perform a fast polarization
transformation at the selected wavelengths. At the same time the
method discussed in this Letter allows one to unambiguously
distinguish all states forming chosen bases. The measurement set-up
which has been already tested in our experiments is shown on the
Fig.~\ref{setup2}.
It consists of the dichroic mirror, separating the photons with the
different wavelenghts, and a pair of polarization beam-splitters,
separating photons with the orthogonal polarizations. We would like
to stress that using dichroic beam-splitter allows one to achieve
100\% mode separation efficiency which is not possible for qutrits.
Four-input double-coincidence scheme linked to the outputs of
single-photon detectors registers the biphotons-ququarts. For
example for the first basis, the scheme works as follows, provided
that Bob's guess of the basis is correct:
\par if state $|H_{1}H_{2}\rangle$ comes, then detectors D4, D2 will fire,
\par if state $|H_{1}V_{2}\rangle$ comes, then detectors D4, D1 will fire,
\par if state $|V_{1}H_{2}\rangle$ comes, then detectors D3, D2 will fire,
\par if state $|V_{1}V_{2}\rangle$ comes, then detectorsD3, D1 will fire.
 Same holds for any of the remaining correctly guessed bases, since
the quarter- and half wave plates transform the polarization to HV
basis in which the measurement is performed. Registered coincidence
count from a certain pair of detectors contributes to corresponding
diagonal component of the measured density matrix. So if the basis
is guessed correctly, then the registered coincidence count
deterministically identifies the input state. We illustrate this
statement by the table which shows total number of registered events
per 30 sec for the input state $|R_{1}L_{2}\rangle$ measured in
circular basis and calculated components of the experimental
(theoretical) density matrix.
\begin{table}[!ht]
\begin{tabular}{|c|c||c|c||c|c||c|c|}\hline
$D_{4}D_{2}$&$\rho_{11}$&$D_4D_1$&$\rho_{22}$&$D_3D_2$&$\rho_{33}$&$D_3D_1$&$\rho_{44}$\\
\hline
$0$&$0.0(0)$&$220$&$0.973(1)$&$6$&$0.027(0)$&$0$&$0.0(0)$\\
\hline
\end{tabular}
\caption{Coincidence rate and density matrix components}
\end{table}
Moreover, registering coincidences allows one to circumvent the
problem of the detection noise that is common for single-photon
based protocols. If the coincidence window is quite small, it is
possible to assure a very low level of accidental coincidences for
the usual dark count rate of single photon detectors.\\To conclude
we have suggested and tested a novel method of the preparation, and
measurement of the subset of four-dimensional polarization quantum
states. Since for this class of states the polarization degree is
not invariant under SU2 transformations it is possible to switch
between states using simple polarization
elements.\\\textbf{Acknowledgments.} This work was supported in part
by Russian Foundation of Basic Research 06-02-16769, MSU
interdisciplinary grant, Russian Agensy of Science and Innovations
2006-RI-19.0/001/593. S.P.K. acknowledges support from RASI 435/2005
.


\begin{thebibliography}{99}
\bibitem{order} The thickness of the retardation plate that performs the transformation is determined by the number of order $m_{i}$ that is chosen
for the specified wavelength according to
$\frac{m_{2}\lambda{2}}{\delta
n_{2}}=\frac{(2m_{1}+1)\lambda{1}}{2\delta n_{1}}$. Here $\delta
n_{i}$ is the value of the birefringence for the specific
wavelength. We chose order $m_{1}=51$ for $\lambda=702nm$ and
$m_{2}=47$ for $\lambda=650nm$.
\end{thebibliography}
\end{document}